# Observation of inherent chiral Smith–Purcell effect via symmetry breaking

Juan-Feng Zhu[1#], Fei Chen[2#], Jun-Xuan Chen[2], Junkang Shan[2], Ayan Nussupbekov[3], Febiana Tjiptoharsono[4], Xuan Dong[1], Mu Wang[2*], Cheng-Wei Qiu[5*], Zhaogang Dong[1,4,6*], Ruwen Peng[2*], Lin Wu[1,3*]

[1]Science, Mathematics, and Technology (SMT), Singapore University of Technology and Design (SUTD); 8 Somapah Road, Singapore 487372, Singapore.

[2]National Laboratory of Sol`23id State Microstructures, School of Physics, and Collaborative Innovation Center of Advanced Microstructures, Nanjing University; Nanjing 210093, China.

[3]Institute of High Performance Computing (IHPC), Agency for Science, Technology, and Research (A*STAR); 1 Fusionopolis Way, #16-16 Connexis, Singapore 138632, Singapore.

[4]Institute of Materials Research and Engineering (IMRE), Agency for Science, Technology, and Research (A*STAR); 2 Fusionopolis Way, #08-03 Innovis, Singapore 138634, Singapore.

[5]Department of Electrical and Computer Engineering, National University of Singapore; 4 Engineering Drive 3, Singapore 117583, Singapore.

[6]Quantum Innovation Centre (Q.InC), Agency for Science, Technology, and Research (A*STAR); 2 Fusionopolis Way, #08-03 Innovis, Singapore 138634, Singapore.

[#]These authors contributed equally to this work.

[*]Corresponding author. muwang@nju.edu.cn (M.W.); chengwei.qiu@nus.edu.sg (C.W.Q.); zhaogang_dong@sutd.edu.sg (Z.D.); rwpeng@nju.edu.cn (R.P.); lin_wu@sutd.edu.sg (L.W.)

The Smith–Purcell effect arises when charged particles move near a periodic structure, emitting radiation. Conventional approaches for generating chiral Smith–Purcell radiation rely on metasurface phase engineering or resonant mode interference, typically producing narrow-band, weakly chiral emission. Here, we introduce a resonance-interference-free mechanism that leverages the properties of the charged particles themselves. Using a non-chiral, non-resonant silicon grating, we demonstrate broadband, tunable Smith–Purcell radiation with high chirality, achieving a record-high degree of polarization of 0.87. This is enabled by converting the transverse spin angular momentum of electron-induced evanescent waves into a longitudinal form, producing opposite chirality at different azimuthal angles. Beam twisting or displacement offers precise control over chirality, paving the way for compact chiral light sources, advanced X-ray imaging, and integrated particle diagnostics platforms.

Accurate particle diagnostics and trajectory control remain major challenges in beam–matter interactions, especially for advanced applications like next-generation linear colliders and free-electron lasers[1]. A conventional approach involves detecting radiation emitted by the beam itself, such as transition[2] or Cherenkov radiation[3]. While effective, these methods are typically invasive, constrained by energy thresholds, and require bulky setups. This highlights the pressing need for non-invasive, cost-effective, and miniaturized diagnostic tools[4]. The Smith–Purcell effect[5] offers a compelling alternative to conventional diagnostics[6–8]: when a charged particle travels near a periodic grating, it emits radiation that is intrinsically linked to its trajectory and energy. However, existing Smith–Purcell effect–based diagnostic schemes are sensitive only to the *magnitude* of the electron trajectory deflection, not its *direction* relative to the grating periodicity. As a result, deflections of opposite signs produce identical spectral responses, limiting the diagnostic information and resolution. Overcoming this limitation would enable real-time beam monitoring and facilitate the seamless integration of diagnostic functionality into compact vacuum electronic devices and on-chip free-electron laser systems.

One promising route to overcoming these limitations lies in leveraging higher-order photonic degrees of freedom—most notably, **chirality**[9]. Introducing chiral properties into Smith–Purcell radiation can greatly enhance its sensitivity, paving the way for high-resolution, non-invasive particle diagnostics. Existing strategies typically rely on metasurface-based phase engineering[10–13] or resonance-induced mode interference[14,15]. For example, electrons injected into a nanosquare light-well can excite interfering modes that yield moderately chiral radiation, with measured chirality up to 0.4[14]. However, these approaches are often mediocre, relying heavily on finely tuned structural parameters, which limit their scalability, robustness, and practical applicability. This raises a fundamental question: **can chiral Smith–Purcell radiation be achieved through a more general, resonance-independent mechanism?**

The answer is **Yes**, and in this work, we demonstrate that chirality is inherently embedded in the Smith-Purcell effect, arising from the spin angular momentum (SAM) of the electron-induced evanescent wave[16–20]. Fundamentally, periodic structures convert the evanescent wave into propagating radiation by providing momentum compensation. During this process, transverse SAM (T-SAM) from the near-field can be converted into longitudinal SAM (L-SAM) in the far-field, which corresponds to circular polarization, revealing the presence of chirality in the emitted light. Our full 3D analysis shows that Smith-Purcell radiation naturally includes left-handed circular polarized (LCP), right-handed circular polarized (RCP), and linearly polarized (LP) components, symmetrically distributed in space. In the symmetry plane—aligned with the electron trajectory—LCP and RCP cancel, yielding predominantly LP radiation, in agreement with traditional 2D interpretations[21]. To induce net chirality, we break this symmetry by shifting or twisting the electron beam relative to the grating. Experimental results confirm this approach enables highly tunable, strongly chiral Smith-Purcell emission. Unlike previous methods that depend on structural resonances and precisely engineered geometries[10–15], our approach overcomes the inherent limitations in bandwidth and chirality control, while preserving the robustness and simplicity of the Smith–Purcell effect. This advancement enables new possibilities for compact, non-invasive particle diagnostics[22], as well as broadband, integrated chiral light sources[23]. Such capabilities hold significant potential for industrial and biomedical imaging applications—for instance, achieving sub-wavelength resolution in X-ray imaging[24] without altering the emission frequency.

## Concept & Working Principle

The fundamental origin of the Smith-Purcell effect arises from the interaction between moving electrons and periodic structures[5]. Consider a single electron moving along the $x$-direction, with the current density $\boldsymbol{J}(\rho,t)=\hat{\boldsymbol{x}}qv\delta(\rho)\delta(x-vt)/2\pi\rho$, where $q$ is the electron charge, $v=\beta c$ is the electron velocity, $\beta$ is the electron velocity normalized by the speed of light $c$, and $\rho=\sqrt{y^2+z^2}$ is the radial distance from the electron trajectory in the transverse plane. Transforming this expression into the frequency domain gives the time-harmonic current density:

$$\boldsymbol{J}(\rho,\omega)=\hat{\boldsymbol{x}}\frac{q}{4\pi^2\rho}e^{\frac{i\omega x}{v}}\delta(\rho). \tag{1}$$

where $\omega$ is the angular frequency. Solving Maxwell's equations results in the electric field[25]:

$$\boldsymbol{E}(\rho)=-\frac{q}{8\pi\omega\varepsilon}\left[\hat{\boldsymbol{x}}k^2+\frac{i\omega}{v}\boldsymbol{\nabla}\right]H_0^{(1)}(k_\rho\rho)e^{\frac{i\omega x}{v}}, \tag{2}$$

where $H_0^{(1)}$ is the zeroth-order Hankel function of the first kind, and $k_\rho$ is the transverse (radial) wavevector component. Since $v<c$, the electron wavenumber ($k_{xe}=\omega/v$) exceeds the vacuum wavenumber ($k_0=\omega/c$ ), causing the induced wave to become evanescent and decay exponentially far from the trajectory (see intensity mapping and the red curve in Fig. 1a). Unlike plane waves, the SAM of electromagnetic fields can be described by a generalized expression, $\boldsymbol{S}=\mathrm{Im}(\varepsilon\boldsymbol{E}^*\times\boldsymbol{E}+\mu\boldsymbol{H}^*\times\boldsymbol{H})/(\varepsilon|\boldsymbol{E}|^2+\mu|\boldsymbol{H}|^2)$, which allows a transverse SAM (T-SAM) component perpendicular to the wavevector in evanescent fields[16–20]. In the present study, the electron propagates in a vacuum, so $\varepsilon=\varepsilon_0$ and $\mu=\mu_0$. Although the electron-induced evanescent field is symmetric about the electron trajectory, the SAM forms a vortex-like distribution around the trajectory (Fig. 1a), with its circulation set by the electron motion. Consequently, opposite SAM orientations appear on either side of the trajectory across a transverse cross-section, as indicated by the blue curve. When the electron passes over a periodic structure, such as a grating, diffraction from the grating compensates for the momentum, converting the electron-induced evanescent wave into a plane wave in the vacuum. This phenomenon is referred to as Smith-Purcell radiation[5], and its corresponding dispersion relation:

$$\lambda=\frac{p}{|m|}\left(\frac{1}{\beta}-\cos\theta\right), \tag{3}$$

where $\lambda$ is the radiation wavelength, $p$ is the grating pitch, $\theta$ is the emission angle relative to the electron's motion, and $m$ is the harmonic order.

While its inherent SAM is expected to be conserved in Smith-Purcell radiation, we verify this by examining a typical Smith-Purcell radiation scenario based on a silicon grating (see Methods and Supplementary Information S1). The parameters are set as follows: pitch $p$ = 120 nm, duty cycle 0.3, depth 150 nm, and normalized electron velocity $\beta$ = 0.3284 (electron energy 30 keV). The SAM distribution at the normal emission angle ($\theta = 90°$) is shown in Fig. 1b. It is observed that $\boldsymbol{S_z}$ maintains the same sign in both near and far fields, indicating the conservation of SAM throughout the emission process. In the far-field, the SAM—referred to as L-SAM because it is aligned with the wavevector—is directly related to the chirality of the radiation[18]. The presence of opposite L-SAM distributions corresponds to opposite circular polarization states in the far field, thereby revealing an inherent chirality in the Smith-Purcell effect. However, this feature has often been overlooked in previous studies[5,10,14,26–28], which typically assume the electric vector of Smith-Purcell radiation is mainly perpendicular to the grating[5]. This assumption is consistent with conventional 2D theoretical analyses conducted at the central $y=0$ plane[21], where the $\boldsymbol{E_y}$ component vanishes, leaving only $\boldsymbol{E_x}$ and $\boldsymbol{E_z}$ components, indicating LP polarization (Figs. S2-S5). This interpretation also aligns with the conservation of SAM, where T-SAM at this interface gives $\boldsymbol{S_z} = 0$, implying zero L-SAM in the far-field, *i.e.*, LP emission (Figs. S6).

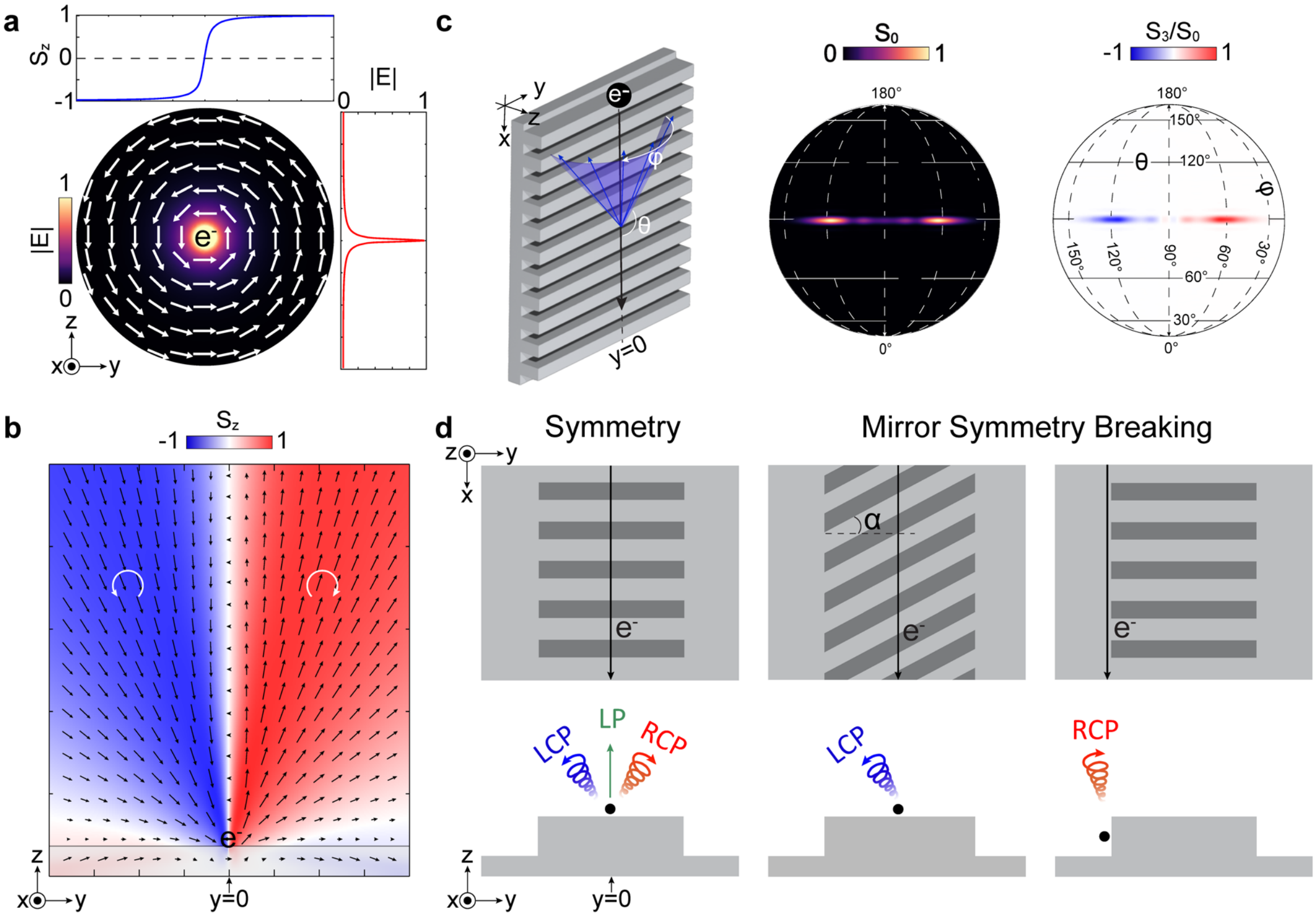

**Fig. 1. Chiral Smith-Purcell Effect**. **a**, Distribution of spin angular momentum (SAM) in the electron-induced evanescent wave, with electric field intensity (pseudocolor) and SAM directions (white arrows). **b**, SAM conservation in the Smith–Purcell effect. Transverse SAM (T-SAM) near the grating is converted to longitudinal SAM (L-SAM) in the far-field. Arrows indicate the SAM vector $\boldsymbol{S}$; the longitudinal component $\boldsymbol{S_z}$, approximately representing far-field chirality, is shown as a pseudocolor map. The shadow region indicates the grating. **c**, **Left**: 3D Smith–Purcell radiation model with the electron moving in the symmetry plane ($y = 0$), exciting a cone-shaped emission pattern: $\theta$ and $\varphi$ denote the polar and azimuthal angles, respectively. **Right**: Stokes parameters distribution in the far-field at $\theta = 90°$, with $S_0$ (intensity) and $S_3$ (chirality). **d**, Inherent chiral Smith–Purcell effect showing central linear polarization (LP), right-handed circular polarization (RCP) on the $+y$ side, and left-handed circular polarization (LCP) on the $-y$ side. Chirality arises from symmetry breaking via electron beam twisting or lateral displacement relative to the grating.

To validate our prediction, we project the field distribution into the far-field (latitude: $\theta$ and longitude: $\varphi$), and the Stokes parameters ($S_0$: intensity and $S_3$: chirality) mappings (See Methods) are shown in Fig. 1c. Interestingly, symmetric intensity ($S_0$) peaks appear at approximately $\varphi = 63°$ and $\varphi = 117°$ rather than at the normal azimuthal angle ($\varphi = 90°$). The chirality distribution ($S_3$) also follows a mirror-symmetric pattern: $S_3 = 0$ at $\varphi = 90°$, with positive (RCP) and negative (LCP) components at $\varphi = 63°$ and $\varphi = 117°$, respectively. Since the intensities of LCP and RCP are equal, the net chirality would be zero (Fig. S7). This is why conventional theory and experiments, even when considering azimuthal angle dependence in a 3D model, readily overlook this key feature[27,28]. Notably, this inherent chirality distribution is universal, independent of the emission angle $\theta$, material composition, or structural parameters. Consistent polarization patterns are observed not only at normal emission ($\theta = 90°$) but also in both forward and backward directions, including for perfect electric conductor (PEC) gratings (Figs. S8-S10).

Although the inherent chirality of Smith-Purcell radiation has been identified, realizing chiral emission requires azimuthally selective collection due to the symmetric distribution of chirality, which inevitably increases experimental complexity. To overcome this challenge, we propose two strategies to break in-plane mirror symmetry: twisting the electron trajectory or laterally displacing the electron beam relative to the grating, as illustrated in Fig. 1d. Throughout this article, all results are based on a simple silicon grating. We will show that this straightforward platform enables broadband, highly chiral, and tunable radiation by harnessing the inherent chirality of the Smith-Purcell effect.

**Symmetry Breaking via Electron-Grating Twisting**

Twisting the electron's trajectory relative to the grating breaks in-plane mirror symmetry, inducing chiral Smith-Purcell radiation. As shown in Fig. 2a, this twist transforms the reciprocal lattice in $k$-space from a rectangle to a parallelogram. Since Smith-Purcell radiation results from the electron dispersion folded above the light cone within the Brillouin zone, its emission properties are intrinsically linked to the underlying symmetry breaking. Projecting the SAM of the evanescent wave into $k$-space reveals an asymmetry: negative SAM dominates for $k_x > 0$, positive for $k_x < 0$, and they roughly balance at $k_x = 0$. The portion of SAM within the deformed light cone determines the radiation chirality, leading to angle-dependent polarization—LCP in the forward direction, RCP in the backward direction, and LP near normal emission. The degree of polarization (DOP), defined as $(I_{\mathrm{RCP}} - I_{\mathrm{LCP}})/(I_{\mathrm{RCP}} + I_{\mathrm{LCP}})$, transitions accordingly, with its slope governed by the twist-induced $k$-space displacement (Figs. S11-S12).

To verify this principle, we conducted full-wave simulations with the same structural parameters (see Methods and Supplementary Information S1). Due to the relative twist angle $\alpha$ between the free electron and the grating, the effective period is extended to $p/\cos\alpha$, and the Smith-Purcell effect dispersion relation is modified as follows:

$$\lambda = \frac{p}{|m| cos\alpha}\left(\frac{1}{\beta} - cos\,\theta\right). \tag{4}$$

The calculated DOP map as a function of twist angle $\alpha$ and radiation wavelength is shown in Fig. 2b. It reveals that LCP (RCP) chiral Smith-Purcell radiation emerges in the forward (backward) direction, while LP is observed near normal incidence, as indicated by the dashed curve. The maximum DOP increases significantly as $\alpha$ rises from 0° to 30°, eventually saturating. At larger twist angles, such as $\alpha = 45°$, near-perfect RCP with a DOP of 0.82 is achieved, as highlighted in the inset of Fig. 2b. Notably, when the radiation angle approaches the emission boundaries ($\theta = 0°$ or 180°), the emission becomes difficult to collect, resulting in a reduced DOP. The simulated $S_3$ distribution in the far-field (Fig. 2c) shows that LCP dominates the –$y$ region in the forward direction ($\theta = 45°$), while RCP appears in the $+y$ region ($\theta = 135°$). At normal incidence, both LCP and RCP are present with distinct spatial separation. The intensity ($S_0$) distribution exhibits a similar spatial pattern, consistent with the conical emission effect[29–31], in which the relative orientation between the electron propagation direction and the lattice reshapes the far-field pattern. Importantly, this geometrical redistribution of intensity does not account for the observed chirality evolution, which arises from symmetry breaking and the associated SAM conversion process.

To experimentally validate the concept, a series of silicon gratings with continuously varying twist angles was fabricated on a single substrate (See Methods and Supplementary Information S4). An SEM image of the grating with $\alpha = 45°$ is shown in Fig. 2a. These gratings were vertically mounted in a scanning electron microscope (SEM), where a focused electron beam moved parallel to their surfaces to excite Smith-Purcell radiation. The emitted radiation was collected by a parabolic mirror and directed to detectors for spectral and polarization-resolved (LCP/RCP) analysis (See Methods and Fig. S17).

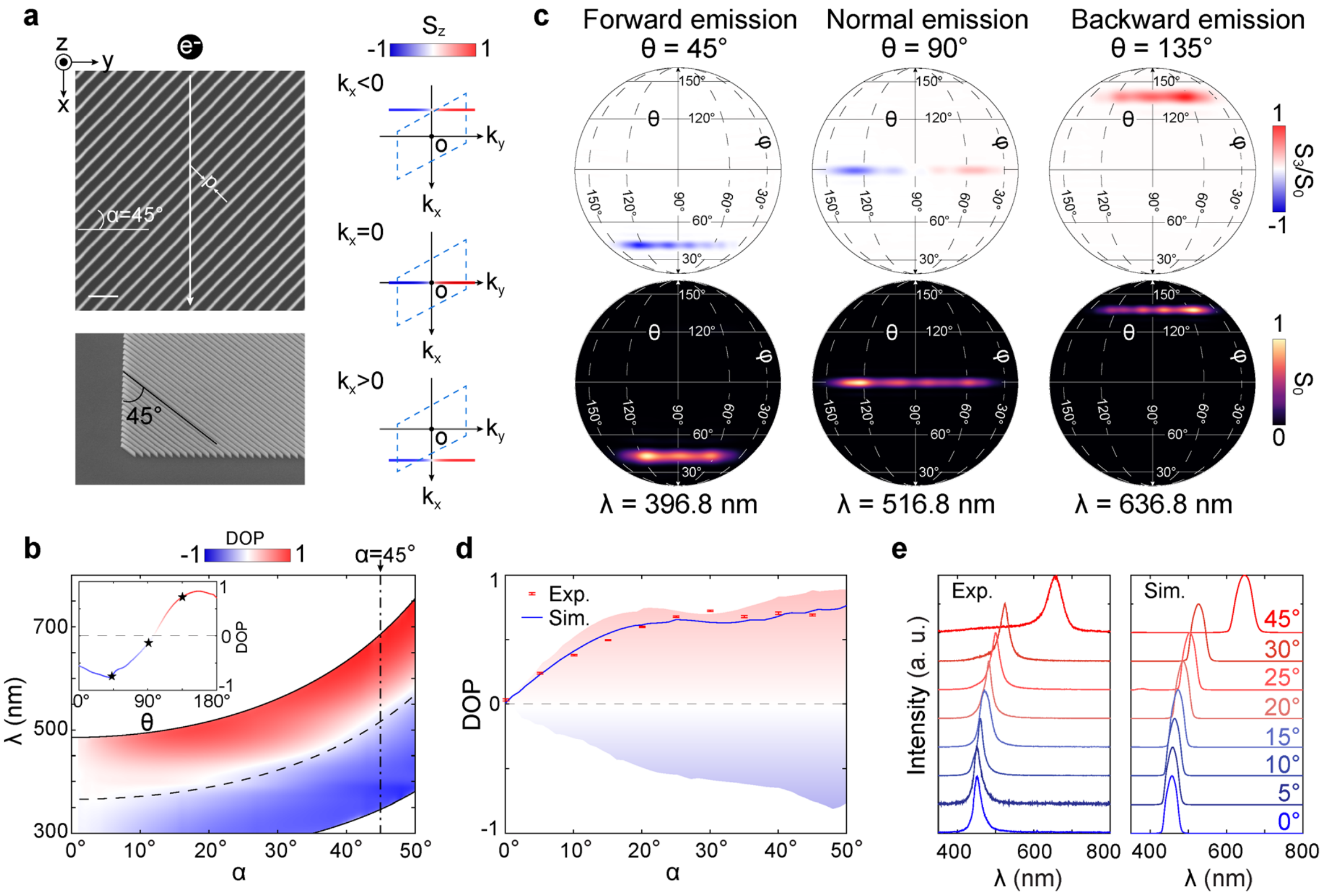

**Fig. 2. Chiral Smith-Purcell radiation via electron-grating twisting**. **a**, SEM image of a grating with a twist angle $\alpha = 45°$ (scale bar: 200 nm) and corresponding $k$-space schematic showing SAM variation within the deformed Brillouin zone. **b**, Simulated DOP vs. twist angle; dashed line indicates normal emission ($\theta = 90°$). Inset: DOP distribution at $\alpha = 45°$. **c**, Far-field $S_3$ and $S_0$ distributions at $\theta = 45°$, 90°, and 135° for $\alpha = 45°$. **d**, DOP vs. twist angle, integrated over emission angles 110°–170°; red bars: experiment, blue curve: simulation, shadow region: simulated DOP for the entire angle range of 0° to 180°. **e**, Measured (left) and simulated (right) emission spectra for varying twist angles; normalized electron velocity $\beta = 0.3284$ (energy 30 keV).

Figure 2d compares experimental and simulated results. The shaded regions represent the simulated DOP range (maximum and minimum) over the entire Smith-Purcell emission spectrum for each twist angle $\alpha$. Red markers denote measured DOP values integrated over the emission angles $\theta$ from 110° to 170°. The blue curve corresponds to the simulated DOP values, integrated over the same angular range as the experiment. The measured and simulated DOP values show strong agreement, with both exhibiting an increasing trend with the twist angle. The measured DOP reaches a maximum of 0.726 at $\alpha = 30°$, while the simulated DOP reaches 0.755 at $\alpha = 50°$. Additionally, the measured emission spectra at different twist angles (Fig. 2e) closely match the simulated spectra, confirming the modified Smith–Purcell dispersion described by Equation (4). The corresponding Stokes parameter distributions at the spectral peaks are provided in Fig. S16.

Importantly, this method is general and not limited by specific grating parameters. To experimentally demonstrate its universality, we varied the grating pitch from 120 nm to 100 nm and achieved a maximum measured DOP of 0.87 — the highest reported to date[14], to our knowledge. The measured average DOP remains above 0.8 across electron energies from 20 keV to 30 keV (Fig. S18), confirming the robustness of high-chirality emission over a broad energy range. Furthermore, reversing the twist angle results in a complete reversal of the emitted chirality (Fig. S19). Simulations further reveal minimal sensitivity to the grating depth (100–160 nm) (Fig. S20a), underscoring the resonance-free nature of the mechanism. We also examined other parameters, such as electron energy, and found that the emission performance remains highly robust (Fig. S20b). Although structurally similar to a prior theoretical twisted-grating study[12], that work relies on sheet-current excitation and structural tuning to generate circular polarization, whereas our point-electron model reveals an inherent chiral Smith–Purcell effect originating from the spin–momentum–locked evanescent field and controllable via symmetry breaking.

**Symmetry Breaking via Electron-Grating Shifting**

Chiral Smith-Purcell radiation can also be achieved by laterally shifting the electron beam, selectively converting one component of T-SAM into L-SAM, and thereby generating chiral emission. From the co-moving frame of the electron (Fig. 3a and top panel of Fig. 3b), shifting the electron toward the right side of the grating (–$y$ region) leads to preferential coupling with the left side of the electron-induced evanescent field (Figs. S21-S23). This asymmetry leads to a net positive $\boldsymbol{S_z}$ in the near-field interaction, resulting in RCP radiation in the far-field (Fig. S24). Unlike the twist-based method, which produces opposite DOP values in the forward and backward directions, this approach yields a consistent chiral state across the full angular range of Smith-Purcell radiation. Additionally, chirality can also be tuned by adjusting the electron's lateral position: shifting the beam to the opposite side (*i.e.*, into the +$y$ region) reverses the handedness of the emitted radiation. The corresponding experimental and simulation results are shown in the bottom panel of Fig. 3b and Fig. S25, respectively.

To examine the transition between chiral and achiral emission, we systematically varied the lateral position of the electron beam relative to the grating. The experimental configuration is shown in the left panel of Fig. 3b, where red dots mark electron-beam positions approximately 20 nm above the grating surface with a step size of ~10 nm. Details of the experimental accuracy, including the electron-beam size, positioning precision, and beam drift, are provided in Supplementary Information S11. During the measurements, the electron-beam position is indicated by a red cursor in the pre-captured SEM image and is scanned from position #0 to #80 with a constant step size. Fig. 3b presents the measured DOP values together with the corresponding LCP (blue) and RCP (red) intensities. As the beam is displaced from position #0 vertically along the sidewall, the DOP initially increases, peaking at ~ 0.52 at position #17. It then decreases due to reduced coupling away from the sidewall and subsequently fluctuates around zero as the beam continues to shift laterally. A similar trend is observed when the electron shifts to the +$y$ side of the grating, where the DOP reaches a minimum of about −0.798 at position #18.

Fluctuations are also evident in the simulated DOP mapping across the full Smith–Purcell radiation range (Fig. 3c), with the simulated spectral peak (Fig. S26) marked by the dashed line. From positions #0 to #17 (corresponding to the side wall region), the electron remains on the $-y$ side of the grating, consistently producing RCP-dominated emission with a high DOP, averaging above 0.8 and peaking at 0.98. For example, the DOP curve at position #9 (inset of Fig. 3c) shows broadband high chirality across the entire emission range. Simulated Stokes parameter mapping at forward, normal, and backward emission angles is shown in Fig. 3d. Since the interaction selectively involves only the positive $\boldsymbol{S_z}$ component (Fig. S24), the resulting far-field intensity exhibits asymmetry, consistent with theoretical expectations.

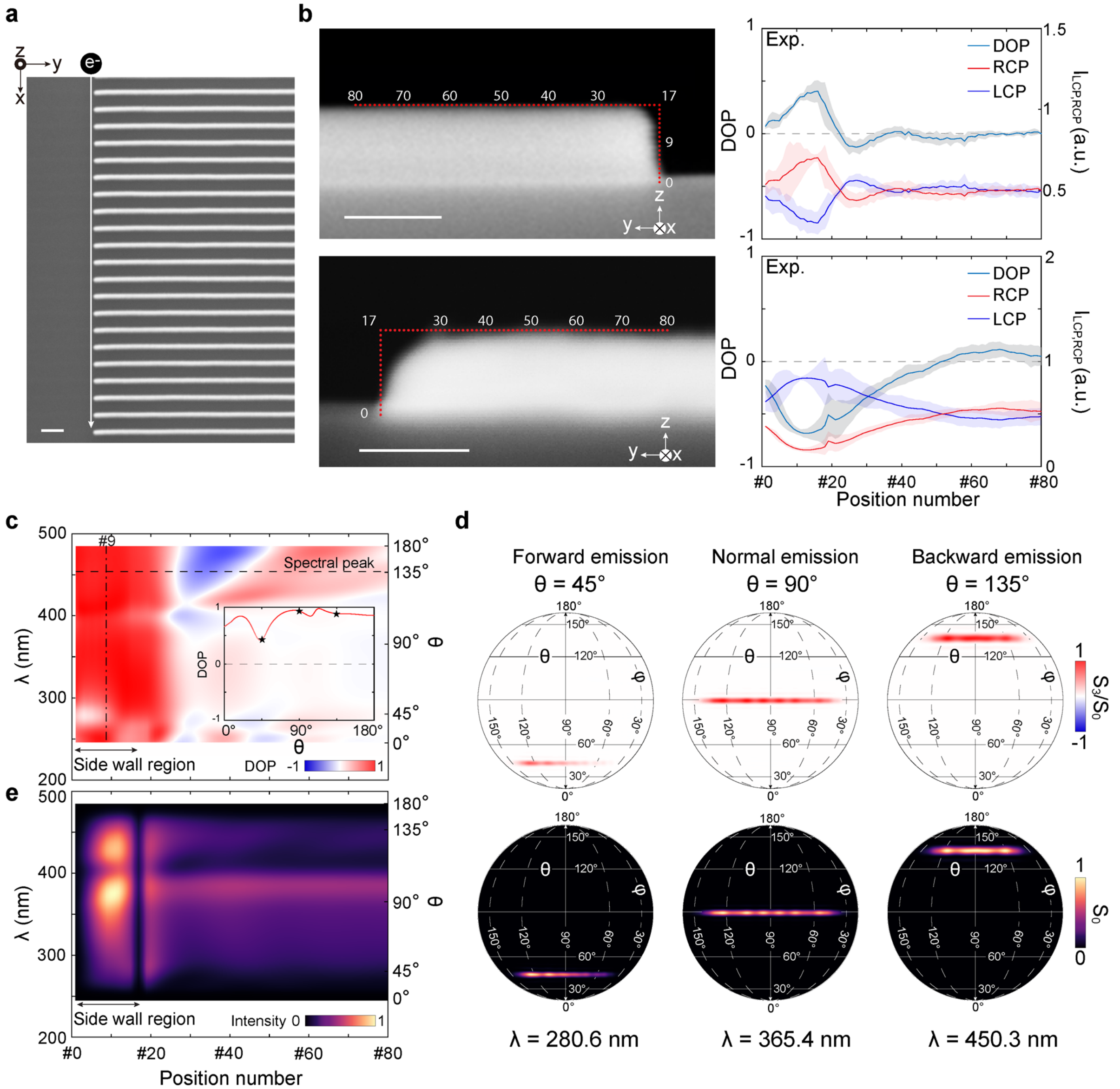

**Fig. 3. Chiral Smith-Purcell Radiation via electron-grating shifting**. **a**, Top-view SEM image of the grating. The electron shifts to the side of the grating rather than moving above the grating surface. (**b**) Experimental results of chiral Smith–Purcell emission when the electron moves laterally along the grating side. The electron shifts to the −y region (top panel) and the +y region (bottom panel). Left: corresponding SEM images, with red dots indicating the incident electron positions. Right: measured DOP, LCP, and RCP as functions of electron position; the shaded regions represent ±1 standard deviation of repeated measurements. (**c**–**e**) Simulated results for the case where the electron shifts to the right side of the grating. (**c**) Simulated DOP mapping as a function of electron position; the inset highlights the DOP variation with emission angle at position #9. (**d**) Simulated Stokes parameters $S_3$ and $S_0$ at emission angles of 45°, 90°, and 135°, corresponding to wavelengths of 280.6 nm, 365.4 nm, and 450.3 nm, respectively. (**e**) Simulated emission intensity as a function of electron position. All scale bars in this figure correspond to 200 nm.

As the electron shifts laterally from position #18 to #80, LCP components are gradually introduced, partially canceling the initial RCP emission and reducing the DOP toward zero, with possible sign reversal (Fig. S26). Between positions #18 and #40, enhanced edge scattering increases the LCP contribution, leading to a transition from positive to negative DOP. Beyond position #40, as the electron moves further from the grating edge, scattering weakens, and the DOP trends back toward zero, resulting in the observed oscillatory behavior. These oscillations are more pronounced at longer wavelengths due to the increased decay length of the evanescent field, especially in the backward emission direction. In contrast, at shorter (blue-shifted) wavelengths, the evanescent field decays more rapidly, reducing edge-scattering effects and suppressing DOP fluctuations.

Interestingly, the strongest emission is observed when the electron is positioned beside the grating, rather than directly above it, as shown in Fig. 3e. This counterintuitive result arises from two key mechanisms: (i) **Corner scattering** – Sharp grating corners act as efficient radiative antennas, coupling electron-induced evanescent waves into free-space radiation with enhanced intensity. Increasing the grating depth and shifting the beam away from the edge significantly reduces emissions, confirming the role of the corners (Fig. S27). This enhancement is also experimentally verified (Fig. S28). Replacing these sharp corners with rounded edges further reduces emissions, further supporting their role (Fig. S29). (ii) **Substrate-induced reflection** – The emission enhancement can be understood in terms of impedance mismatch at the substrate interface. A substrate with a large complex refractive index ($n+ik$) has an electromagnetic impedance significantly different from that of free space, resulting in strong reflection. Consequently, the electron-induced evanescent field is reflected toward the grating rather than propagating into the substrate, suppressing internal radiation channels[32] and redirecting electromagnetic energy outward, thereby enhancing the observable Smith–Purcell emission (Figs. S30–S31). This behavior challenges the conventional assumption that material loss necessarily reduces radiation efficiency. Additionally, emission intensity is highly sensitive to the electron–substrate distance, decaying rapidly as the electron approaches or moves away from the optimal lateral position. At the farthest measured point along the electron's curved trajectory (position #17), where only a small portion of the evanescent field interacts with the grating, the emission is minimal.

Importantly, this underlying mechanism imparts strong robustness to the chiral Smith-Purcell radiation. To verify this, we varied key structural parameters, including grating depth and period, as well as the electron beam energy. In all cases, a high DOP was consistently achieved when the electron was positioned near the side of the grating (Fig. S32).

## Discussions

We demonstrate the robust and flexible generation of chiral Smith-Purcell radiation from a non-chiral periodic structure, independent of structural resonances or fine-tuned parameters. This opens new avenues for applications leveraging the chiral Smith-Purcell effect across a broad spectral range.

Notably, our approach enables integrated chiral light sources with greatly relaxed design constraints. Unlike widely used optical approaches, such as chiral bound states in the continuum[23], where chirality arises from symmetry-broken resonant modes and the emission wavelength, angle, and DOP are highly sensitive to structural details; our method relies only on the grating periodicity to provide momentum compensation and satisfy the Smith–Purcell phase-matching condition. Consequently, the emission wavelength can be flexibly tuned via the grating pitch and electron energy, independent of any structural resonance, enabling substantially greater design and implementation flexibility.

Although the proposed mechanism is universal and applies to both dielectric and metallic gratings, practical experiments require careful material selection due to potential background luminescence. Under electron-beam excitation, materials can produce cathodoluminescence or defect-related photoluminescence that spectrally overlaps with Smith–Purcell radiation, complicating spectral interpretation. For example, $SiO_2$ exhibits visible range cathodoluminescence under electron irradiation. Experiments on gratings fabricated on $SiO_2$ substrates (Fig. S33) show that while the Smith–Purcell peak shifts with electron energy, the cathodoluminescence remains nearly fixed, creating background signals that interfere with reliable detection. These results underscore the importance of choosing appropriate materials in Smith–Purcell emission experiments.

Additionally, the broadband excitation provided by free electrons eliminates the need for external seed light sources and enables chiral emission across an exceptionally wide spectral range—from radio frequencies to X-rays[33,34]—as governed by Eqs. (3) and (4). In this work, we demonstrate first-order broadband Smith–Purcell radiation in the visible regime, with wavelengths down to ~245 nm; extending the inherent chiral Smith–Purcell effect to higher diffraction orders could further access the deep-ultraviolet range[33], enabling higher-resolution nanofabrication and offering a route toward coherent, super-radiant chiral sources through synchronization of electron bunching and harmonic emission[35,36]. Moreover, increasing the electron energy toward the relativistic regime or reducing the grating pitch to the atomic scale—potentially using van der Waals materials—could, in principle, extend chiral Smith–Purcell radiation into the X-ray regime, providing a pathway toward chiral X-ray sources and new contrast mechanisms for X-ray imaging and spectroscopy[24,37]. Finally, the use of all-silicon gratings combined with on-chip electron sources supports compact, efficient, and scalable integration of chiral light sources across this broad spectral range[38–42].

Beyond light generation, this approach holds promise for **miniaturized**, **multi-dimensional particle detectors**[8,43,44]. Chirality-sensitive radiation detection could enable precise, non-invasive particle trajectory recovery and system calibration, complementing conventional invasive methods. Finally, our findings highlight unexplored aspects of chirality in free-electron radiation, potentially inspiring further investigation into related phenomena[45–49] such as transition[2], and Cherenkov[3] radiation, as well as interactions with time-varying media[50–52]. More broadly, the underlying physics provides a generalizable framework for tunable chiral emission in other evanescent-wave systems[19,53], including surface plasmon polaritons[54] and photonic skyrmion[55], through controlled transformation between T-SAM and L-SAM.

## Methods

### Full-Wave Simulations

The full-wave simulations are performed using commercial Ansys Lumerical FDTD (Finite-Difference Time-Domain) software. The simulation domain consists of two distinct regions: a silicon grating and the surrounding vacuum. A single electron, positioned 20 nm above the surface of the silicon grating, excites the Smith-Purcell radiation. The refractive index of silicon is taken from the dataset provided by Palik for the material's optical properties. The induced current density due to the moving electron is described by the equation: $\boldsymbol{J}(\rho, t) = \hat{\boldsymbol{x}} q v \delta(\rho)\delta(x - vt)/(2\pi\rho)$, where $q$ is the electron charge, $v$ is the electron velocity, and $\rho = \sqrt{y^2 + z^2}$ is the radial distance from the electron trajectory in the transverse plane. $\delta(\rho)$ and $\delta(x - vt)$ are Dirac delta functions ensuring the electron's position is localized in space and time, respectively. By applying a Fourier transform, the frequency-domain expression for the current density becomes: $\boldsymbol{J}(\rho, \omega) = \hat{\boldsymbol{x}} q e^{i\omega x/v} \delta(\rho)/(4\pi^2\rho)$. In the simulation, a dipole array polarized along the direction of the electron's motion is employed to model the dynamics of a moving electron, as both share the same electric field profile. The dipole spacing is set to $\Delta x$ =10 nm, well within the subwavelength regime relative to the simulated wavelength range (300–900 nm). A phase delay of $\Delta x\omega/v$ is introduced between adjacent dipoles to emulate the continuous motion of an electron traveling across the grating surface. The simulation results, including dynamic visualizations, are presented in Supplementary Information S1.

### Chirality Analysis

To analyze the distribution of circularly polarized states in Smith-Purcell radiation, a surface monitor is placed above the grating to collect the emitted radiation. The electric field is projected onto a hemisphere to analyze the far-field emission and the far-field electric field components, $\boldsymbol{E_r}$, $\boldsymbol{E_\theta}$, and $\boldsymbol{E_\varphi}$, are extracted. The Stokes parameters are then calculated to characterize the polarization properties of the radiation. These Stokes parameters are defined as follows: $S_0 = \boldsymbol{E_\theta}\boldsymbol{E_\theta^*} + \boldsymbol{E_\varphi}\boldsymbol{E_\varphi^*}$, $S_3 = -2Im(\boldsymbol{E_\theta}\boldsymbol{E_\varphi^*})$, where $S_0$ represents the total intensity (or power) of the radiation, and $S_3$ quantifies the chirality of the emitted field. The value of $S_3/S_0$ = 1 or −1 indicates perfect right-handed circular polarization or left-handed circular polarization, respectively. The collected intensity data is normalized relative to the excitation waveform to ensure that the results are independent of the excitation range, enabling a consistent comparison of polarization properties across different simulation conditions. Additionally, to construct the full 3D radiation pattern, emission data is gathered from six monitors placed strategically around the simulation domain. This allows for a comprehensive analysis of the radiation emitted from the grating.

### Fabrication

Hydrogen silsesquioxane (HSQ) etching masks were fabricated using a 30-nm-thick HSQ layer, which was spin-coated onto a silicon substrate at 5000 rpm using Dow Corning XR-1541-002 resist. The layout for the silicon gratings was generated using MATLAB. Electron beam lithography (Elionix) was performed with a 100 keV acceleration voltage, 500 pA beam current, and an exposure dose of approximately 12 mC/cm². After exposure, the sample was developed in a NaOH/NaCl saline solution (1% wt./4% wt. in deionized water) for 1 minute, followed by a 1-minute rinse in deionized water[56]. The sample was then rinsed with acetone and isopropanol (IPA) and dried with a continuous nitrogen flow. Finally, silicon etching was conducted using inductively coupled plasma (ICP, Oxford Instruments Plasma Lab System 100) with $Cl_2$ gas at a flow rate of 22 sccm, under a process pressure of 5 mTorr, an RF power of 100 W, and ICP power of 150 W, at 40 °C[57].

## Optical Characterization

The sample was vertically aligned in a field-emission scanning electron microscope (SEM, Zeiss ULTRA 55) for Smith-Purcell radiation (Supplementary Information S5). The sample stage allowed for precise rotation and tilting, ensuring the electron beam traveled parallel to the sample surface. Smith-Purcell radiation was excited by the electron beam passing over the surface, collected by a parabolic mirror, and characterized using a cathodoluminescence (CL) detector system (Gatan Mono CL4). The electron beam's accelerating voltage was adjustable between 1 kV and 30 kV, depending on the experimental requirements. The Smith-Purcell radiation excitation spot was positioned at the focal point of the parabolic mirror to maximize collection efficiency. The Smith-Purcell radiation with emission angles ranging from approximately 110° to 170° was reflected by the mirror and directed toward a grating spectrometer equipped with a CCD camera for spectral analysis. Alternatively, the radiation was directed to a highly sensitive photomultiplier tube (HSPMT, 160-930 nm). A left-handed or right-handed circular polarizer (Edmund Optics, CP42HE and CP42HER, 400-700 nm) was placed in front of the photomultiplier tube to measure the left-handed or right-handed circular polarization intensity, respectively.

### Acknowledgments

We thank Dr. W.J. Zhou, Mr. D. Gromyko, Mr. G.X. Liu, Dr. F.Z. Shu, and Dr. C. Wang from Singapore University of Technology and Design; Dr. Z.C. Song from Harbin Institute of Technology; Dr. J.F. Chen from the National University of Singapore, Prof. R.M. Liu from Henan University, and Dr. Z.W. Zhang, Ms. Y.L. Lei, and Prof. C.H. Du from Peking University for their valuable contributions to insightful discussions and assistance with the preparation of the manuscript.

### Funding

This work was supported by the National Research Foundation Singapore (NRF-CRP26-2021-0004, NRF-CRP31-0007), the Ministry of Education Singapore (MOE-T2EP50223-0001, MOE-MOET32024-0005, MOE-T2EP50125-0018), Agency for Science, Technology, and Research (A*STAR) (MTC IRG M24N7c0083), and the Singapore University of Technology and Design (SUTD) (Kickstarter Initiatives SKI 2021-02-14). M.W. and R.P. acknowledge the support from the National Key R&D Program of China (Grant Nos. 2020YFA0211300 and 2022YFA1404303) and the National Natural Science Foundation of China (Grant No. 12234010). Z.D. acknowledges support from A*STAR (MTC IRG M22K2c0088), the National Research Foundation (NRF-CRP30-2023-0003), and the SUTD (Kickstarter Initiative SKI 2021-06-05). C.W.Q. acknowledges support from the Ministry of Education in Singapore (grant nos. A-8002152-00-00 and A-8002458-00-00).

### Author contributions

J.F.Z., C.W.Q., and L.W. conceived the idea. J.F.Z. and A.N. developed the theoretical framework and performed numerical simulations. F.T., X. D., and Z.D. fabricated the samples. F.C., J.X.C., J.S., M.W., and R.P. performed the optical measurements. J.F.Z. and F.C. analyzed the data. All authors contributed to the discussion and interpretation of the results. J.F.Z. and L.W. wrote the manuscript with input from all authors. M.W., C.W.Q., Z.D., R.P., and L.W. supervised the project.

### Competing interests

The authors declare that they have no competing interests.

### Data and materials availability

The data supporting the findings of this study are available in the article and the Supplementary Information.